# Computational Science and Innovation

D. J. Dean[1]

Physics Division
Oak Ridge National Laboratory
Oak Ridge, TN 37831-6373

E-mail: David.Dean@science.doe.gov

**Abstract**. Simulations – utilizing computers to solve complicated science and engineering problems – are a key ingredient of modern science. The U.S. Department of Energy (DOE) is a world leader in the development of high-performance computing (HPC), the development of applied math and algorithms that utilize the full potential of HPC platforms, and the application of computing to science and engineering problems. An interesting general question is whether the DOE can strategically utilize its capability in simulations to advance innovation more broadly. In this article, I will argue that this is certainly possible.

## 1. Introduction: Policy and competitiveness

In this article, I will explore how the U.S. Department of Energy (DOE) can utilize one of its core capabilities – simulations – in order to positively affect U.S. innovation and competitiveness. I begin by pointing out some of the policy drivers for science and innovation that have implications for computing broadly. I will then discuss some of the current capabilities and challenges to continued progress in computational science, and some of the exciting science that lies ahead in nuclear physics. I will then describe how one might apply the DOE's computational science capability to accelerate the innovation cycle in the U.S.

President Obama has said of the United States that "we have a choice to make. We can remain one of the world's leading importers of foreign oil, or we can make the investments that would allow us to become the world's leading exporter of renewable energy. We can let climate change continue to go unchecked, or we can help stop it. We can let the jobs of tomorrow be created abroad, or we can create those jobs right here in America and lay the foundation for lasting prosperity". The President has also said that "[W]hether it is improving our health or harnessing clean energy, protecting our security or succeeding in the global economy, our future depends on reaffirming America's role as the world's engine of scientific discovery and technological innovation".

The world relies on non-renewable fossil fuels to power the economy and the burning fossil fuels produces wealth along with $CO_2$ and other green house gases. A representation of this relationship can be seen in Figure 1 where the per-capita $CO_2$ production is plotted as a function of per capita GDP for various nations. While many lessons can be drawn from this graph, two are important here: first, the curves always move up and to the right. Development entails an increased use of fossil energy (except in the case of France where nuclear power has become a major source of energy). Second, China is rapidly growing. In fact, China has surpassed the U.S. in both energy use and $CO_2$ production in the last year, although its per capita numbers remain well below the U.S. The important point here is that the world will continue to need energy if world-wide standards of living are going to continue to increase.

**Figure 1: $CO_2$ production vs. GDP per capita [1].**

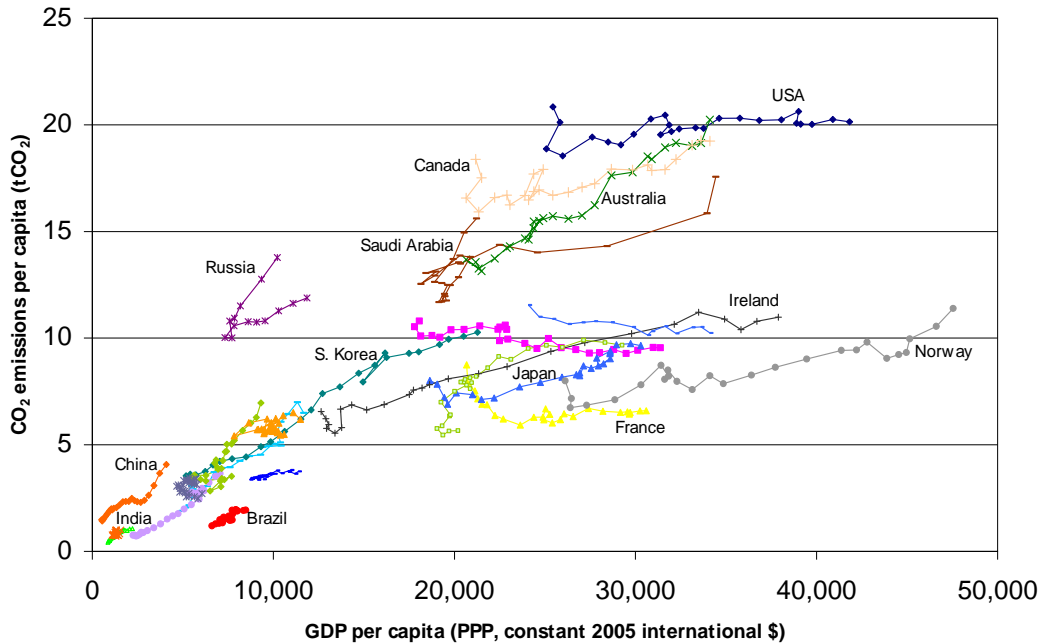

The global climate is affected by the total amount of $CO_2$ and other green house gases in the atmosphere. Green house gases do not absorb or scatter visible light coming from the sun, but they do absorb and re-scatter infrared radiation coming from the earth's surface. Continuous measurements at Mauna Loa Observatory indicate that $CO_2$ levels are rising globally at a rate of about 1.5 ppm per year. The global average atmospheric $CO_2$ content now stands at about 390 ppm, as compared to a pre-industrial concentration of about 280 ppm. Evidence suggests that other GHGs, such as methane, are also increasing in the atmosphere. The overall effect, due to the absorption spectrum of these gases, should be an increase of trapped heat in the atmosphere. Thus, the unconstrained burning of fossil fuels is having an impact on climate change. Figure 2 shows the global warming occurring over the last century relative to the mean temperature from 1951-1980. Predictions of substantial warming over the next 90 years will have a serious impact on society, and could lead to global instabilities if water supplies are threatened or if a substantial rise in sea levels displaces populations [2]. The world is an amazingly well-balanced machine, and while feedbacks are not completely understood in the climate system, there should be cause for concern and planning for adaptation and mitigation [3].

The U.S. goals to reduce dependence on fossil fuels and to decrease GHG emissions will be hard to meet without consistent policies aimed at influencing the national appetite for energy (e.g. CAFE standards for automobile fuel economy) and a consistent effort by the U.S. to identify and fund the research and development that would enable a significant market penetration of green energy technology. For example, the technologies considered in Ref. [4] included advanced fossil fuel liquids, biomass energy, carbon capture and sequestration, efficient electricity generation and distribution, electric drive vehicles, energy-efficient buildings, energy-efficient industrial practices, energy-efficient transportation, nuclear energy, solar energy, and wind energy. The study indicated that meeting both energy goals requires a high probability of success (>50%) for all technologies considered. This finding, coupled with the added complexity of the energy system, particularly on the supply side where a vast infrastructure is already in place, should serve as a sobering reminder that the problems we are facing in energy are extremely difficult.

The Department of Energy is one of the primary parts of the U.S. government concerned specifically with energy, and the missions of the Department align well with the vision that the President has articulated. These missions are to sustain a world-leading basic research capability, both discovery and mission driven; to catalyze a transformation of the national and global energy system; to enhance nuclear security; and to contribute to U.S. competitiveness and jobs. For the remainder of this article, I will explore how simulations can play a role in enhancing U.S. competitiveness.

The argument I will construct is in the form of an analogy. During the last few years, nuclear physicists have embraced the idea that simulations can be used in conjunction with experiments to move discovery forward. In a similar way, utilizing validated simulations in the energy and technology sectors should enable one to accelerate the innovation cycle. I begin with a description of current Department computational capabilities and plans for the future.

**Figure 2: Global temperature anomaly relative to the 1951-1980 mean for annual and 5-year running means through 2009. Green bars are 2σ error estimates [5].**

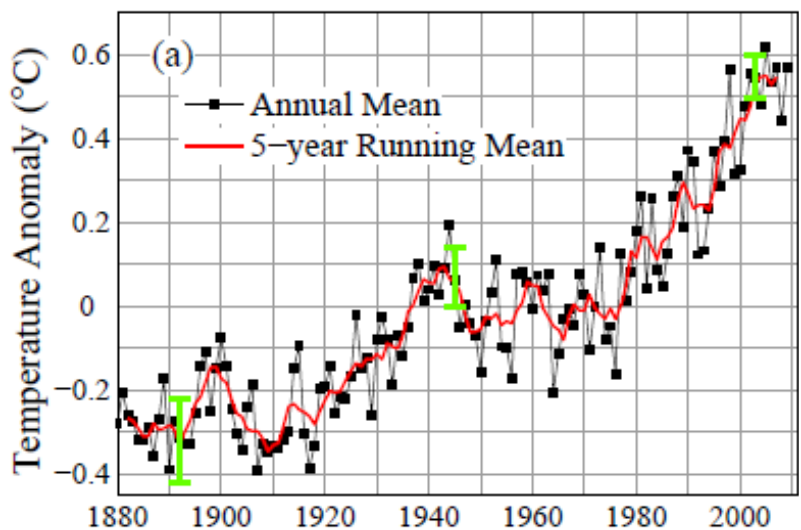

## 2. Simulations at the Department of Energy

Over the years, the Department has developed an impressive set of computational assets with which to tackle problems. For example, in 1992 the U.S. stopped underground nuclear testing, and a new approach was needed to certify the safety, security, and effectiveness of nuclear weapons. This new approach required the development of computer software and hardware well beyond those available at the time and novel experimental capabilities to investigate the most extreme states of matter, previously only accessible in nuclear weapon explosions or in astrophysical objects such as the cores of giant planets and supernovae. Over the next decade, the Stockpile Stewardship Program would transform the capabilities used to understand nuclear weapons performance. As a part of the SSP, the Advanced Simulation and Computing Initiative drove the development of computers that were 10,000 times more powerful and simulation codes that operated in three dimensions with high fidelity representations of the underlying physical phenomena.

In the same time frame, within the Office of Science at the DOE, the Office of Advanced Scientific Computing Research (ASCR) has also pursued computing in open scientific applications. The ASCR program encompasses facilities, applied mathematics, computer science, and the Scientific Discovery through Advanced Computing (SciDAC) program. The SciDAC program was launched in 2001 to develop new tools and techniques for advancing domain-specific scientific research through computational modeling and simulation in all mission areas within the Office of Science and in other areas of the Department.

The DOE computational facilities have become intricate and large, housing the world's fastest and most power-hungry computers. The facilities would not be useful were it not for other aspects of computational science generally. For example, applied mathematics researchers develop the methods that translate the physical world into a language computers understand. They analyze how models behave, optimize algorithms, and explore codes to meet new challenges. Furthermore, computer scientists develop system-level software that integrates and makes usable the HPC platforms on which codes run.

The facilities aspect of what the Department has accomplished in computing over the years cannot be overstated. For example, Department computers have dominated the Top500 list for years. I show in Table 1 the current top computers at various national laboratories throughout the DOE complex. The speeds of these platforms, measured in petaflops ($10^{15}$ floating point operations per second) were determined by a LINPACK suite of test runs as given in the third column. Another interesting characteristic of these machines is their power consumption. For example, the number 1 machine on the Top500 list, Jaguar, requires 6.9 MW to operate.

**Table 1: Top computers at DOE Laboratories (taken from the June, 2010 Top500 list [6]). The speed refers to the Linpack Benchmark performance of the machine.**

| Machine | Place | Speed (Max) | On list since (place) |
|---|---|---|---|
| **Jaguar** | ORNL | 1.75 PF | 2009 (1) |
| **Roadrunner** | LANL | 1.04 PF | 2009 (3) |
| **Dawn** | LLNL | 0.478 PF | 2007 (8) |
| **BG/P** | ANL | 0.458 PF | 2007 (9) |
| **Red Sky (NREL)** | SNL | 0.434 PF | 2010 (10) |
| **Red Storm** | LLNL | 0.416 PF | 2009 (12) |
| **NERSC** | LBNL | 0.266 PF | 2008 (18) |

One of the interesting things about computational capability is that it refreshes quickly. The standard cycle for new chip design is about two years, and HPC machines typically have a 3-5 year lifetime. Until recently, Moore's law [7] drove underlying computational capability through increasing clock speeds and decreasing characteristic component sizes on a chip; however, this trend has now been supplanted by a move to multi-core architectures on a single chip with multi-chips on a single node; Jaguar at ORNL represents this architecture. The other innovation that has occurred in the last few years is a move to heterogeneous architectures where a combination of CPUs and graphical processing units (GPUs) is utilized to enhance parallelism on a node; Roadrunner at LANL is a representative of this architecture. These new and more complicated hardware configurations lead to a need to design underlying algorithms and integrated codes so that they can effectively utilize the parallelisms inherent in the hardware.

The computational hardware tools have to be constantly reinvigorated, and what is today's high-end computer may well be tomorrow's desktop. This can lead to vigorous competition in HPC. For example, the current number 2 machine on the Top500 list is located in Shenzhen, China. The Chinese system, called Nebulae, is constructed from U.S. chips and a 3-layered Infiniband interconnect [8] and achieved 1.27 PF running the Linpack benchmark. The Chinese are expected to surpass the U.S. when the next Top500 list appears in November, 2010. The general trends are shown in Figure 3. While competition is a welcome, and indeed necessary market force, the U.S. cannot afford to rest in its pursuit of HPC. HPC also feeds itself as the machines of today are used to design the chips of tomorrow.

An ambitious target for the next generation of HPC will be platforms that are 1,000 times more powerful than today's petascale platforms. The first question one should ask is whether such a hardware advance is necessary. The justification for extreme scale computing into the next decade was laid out in a series of workshop reports [9] that covered climate, basic energy science, national security, biology, high-energy physics, nuclear physics, fusion energy, and nuclear energy. Significant progress in these areas has been made up to the petascale level, but in each area, significant scientific and engineering challenges were identified that would require the application of capability beyond current reach. One area that has direct implications of the national energy picture is in materials design. The Edisonian approach to design of new materials by trial and error in the laboratory can be vastly accelerated by utilizing simulation to indicate what molecules should be tried that would produce a desired property. This kind of 'materials by design' capability has long been a quest in materials science and should be realized with new computational capabilities currently being envisioned.

In order to achieve an exascale ($10^{18}$ flops, 1,000 times petascale) capability, several R&D challenges will have to be overcome. A first class constraint will be on power. No one is going to buy a gigawatt power plant in order to run an exascale computer. Future machines will require significant innovation for system power reductions, and industry is working toward designs that will enable a more capable chip that requires less power. Related to the power issue is also the memory issue. One of the major areas of innovation will be in memory and data movement. The goal will be to reduce the 100 picojoules/bit that it takes today to move data in and out of main memory, down to 10 picojoules/bit. Programming languages and software capability will also need to be developed. Systems software will have to keep a machine running that has as many as 100 million cores and accelerators. New algorithms that enable concurrency across such a large distribution of cores and accelerators will need to be invented. Finally, the scientific codes will have to be adapted to new architectures and will have to be developed in conjunction with the development of the computers. With careful planning and execution, it will be possible to deliver an exascale platform by the end of the decade.

**Figure 3: China and U.S. top computers in the Top500 list.**

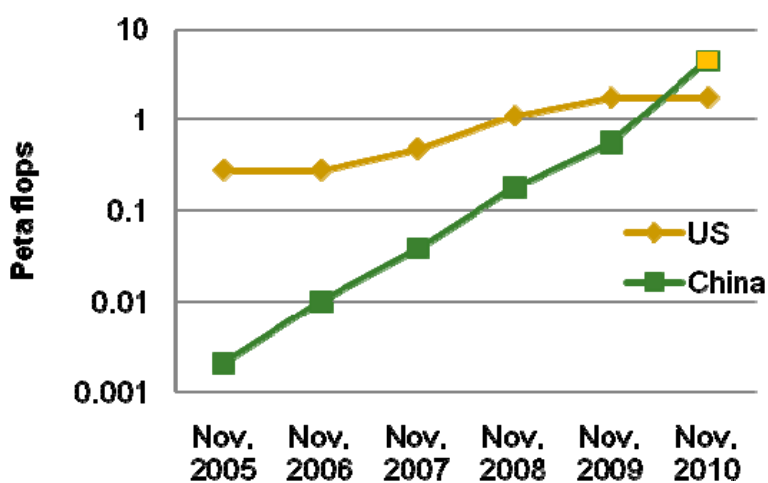

## 3. The Analogy: Nuclear physics and simulations science

In the U.S., nuclear physicists seek to understand how QCD operates in a hot environment, and the underlying quark and gluon structure of the nucleon in a cold environment. Nuclear physicists strive to understand and create new nuclei with exotic properties, and how to use nuclei to detect other exotic particles such as neutrinos and dark matter. And nuclear scientists, in collaboration with scientists from other fields, seek to apply the knowledge gained through experimental and theoretical means to problems in astrophysics, standard model physics, quantum many-body physics, national security, and nuclear energy. This broad program of research is funded through DOE and also through partnership with the National Science Foundation (NSF).

The tools of nuclear physics include DOE national user facilities: RHIC at Brookhaven National Lab [10], CEBAF at Thomas Jefferson National Accelerator Facility [11], HRIBF at Oak Ridge National Lab [12], and ATLAS at Argonne National Lab [13]. In addition, the NSF stewards the NSCL at Michigan State University [14]. The NSF and DOE also maintain facilities at Florida State, Texas A&M, Yale, and Lawrence Berkeley Laboratory. Through these experimental facilities, one may uncover new properties of nuclei and their constituents, and may also enable interesting discoveries for applications in other areas.

These facilities also enable one to validate theory and simulation of the physical systems being experimentally probed. The great tenant of physics – falsification – requires that theory and experiment move hand in hand in a challenging mode. The goal, whether one discusses nuclear physics or climate modelling, is to provide the best possible description of a physical system utilizing simulation as one of the important tools. The tools of simulation extend far beyond just the computational hardware and have a role for theoretical physics that is similar to the role of a national user facility for an experimentalist.

The computational tools have to be applied to physical problems that will have an impact on scientific discovery in a community. Application of high-performance computing to the nuclear quantum many-body problem has enabled significant progress in the last few years, and in the following paragraphs, I will discuss some recent advances in this domain. One can approach the quantum many-body problem in a number of ways but each is associated with solving Schrödinger's equation, or an approximation to it. In nuclear physics, the most direct path toward these solutions is the application of a quantum Monte Carlo technique to the many-body Schrödinger equation in real space. Calculations of this sort have been performed using Greens Function Monte Carlo (GFMC) techniques developed over the years [15]. The GFMC approach is limited to local interactions of phenomenological derivation, but it can explain significant amounts of data in light nuclei through $^{12}$C, provided that the two-body nuclear force is supplemented by a three-body nuclear force.

The nuclear interaction may also be derived from a low-energy expansion of this force utilizing methods of effective field theory (EFT). This approach yields a consistent two- and three-body nuclear interaction [16]. In either the GFMC derivation or the EFT derivation, the parameters of the two- and three-body nuclear forces have to be taken from experimental scattering and bound-state data in the two- and three-nucleon systems. Once determined, these forces can then be used to calculate a variety of physical observables such as ground-state masses, excitation spectra, and transition matrix elements. For example, the GFMC forces were recently applied to an interesting problem of electromagnetic transitions in $^{10}$Be where it was found that the three-body force has a substantial effect on the transition strength [17].

In a similar manner, forces derived using effective field theory methods may also be used to confront experimental data. In the EFT case, however, different techniques are used to solve the quantum many-body problem. One of these is called the No-Core Shell Model (NCSM) [18]. NCSM is a basis expansion technique. In principle, the basis (usually taken as a harmonic oscillator) can be expanded such that one obtains a specified convergence for the problem. The method scales as a combinatorial of the number of basis states and nucleons, and is therefore limited to light nuclei. However, it is invaluable in determining the character of the nuclear interactions, particularly when derived through the EFT expansion methods [19]. Comparison to data then determines whether there is a need to further adjust parameters or whether there is something incomplete in the description of the nuclear interaction.

Coupled-cluster techniques have been used to describe closed- and select open-shell nuclei through [20] mass 48 using EFT-derived two-body forces and have provided forays into three-body applications [21]. The coupled-cluster approach utilizes the idea of an expansion of the many-body wave function in terms of an exponential of low-order one-, two-, (and three-) body excitation operators. For three-body clusters, the method scales like $N^7$, where N is the number of single-particle orbitals in the calculation. Furthermore, the method only allows for linked cluster diagrams, and is therefore size extensive. This can be compared to the diagonalization techniques which expand the many-body wave function in terms of a linear excitation operator and therefore suffer from problems of size extensivity if one truncates the excitation operator series. The challenge is to push the coupled-cluster technique to heavy nuclei with the three-body force included.

The nuclear many-body problem can also be attacked through the use of density functional theory (DFT). Efforts to incorporate an effective field theory treatment into the DFT formalism show promise for establishing the nuclear DFT on a firm theoretical foundation [22]. Typically, the parameters of the functional are usually found by a fitting procedure across many known nuclei (or specially selected nuclei), and comparisons are made to experimental data. These approaches have shown continuing progress over the years as further data has been incorporated into the fitting procedures [23, 24].

In all of these problems, the necessary ingredient for continued success is collaborations with applied mathematicians and computer scientists to create codes and algorithms that will scale to the largest machines in order to effectively utilize computational resources at the highest end to produce science. For example, applications of GFMC to the mass 12 systems required a significant effort through the SciDAC program to enable scaling of the code to more than 100,000 cores. Scientists worked in collaboration with computer scientists at Argonne National Laboratory to produce an Asynchronous Dynamic Load-Balancing Library to apply to the problem [25]. A similar story can be told for the NCSM technique [26] where applied mathematicians worked with domain scientists to achieve scale up to the largest computer systems available. Coupled-cluster techniques are also benefiting from similar interactions [27].

**Figure 4: The use of computation in nuclear physics**

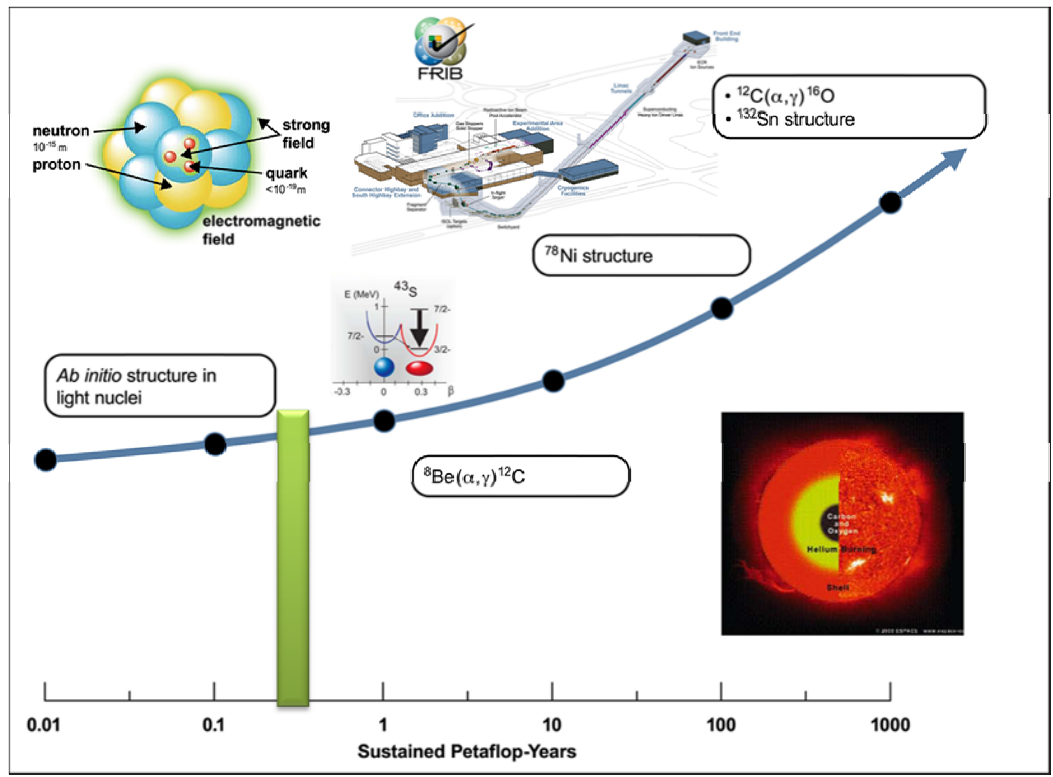

Science proceeds through development of new techniques and capabilities that enable one to predict and measure. One of the recent extreme scale computing workshops was dedicated to understanding how future computing capabilities will impact nuclear physics. The workshop for nuclear physics was hosted by the Office of Nuclear Physics and the Office of Advanced Scientific Computing Research at DOE [28] and covered research in nuclear astrophysics, nuclear structure and reactions, and QCD. In the area of nuclear structure and reactions, the focus was on understanding reactions important in stellar environments, understanding the nature of weakly bound nuclei, probing neutrino physics with nuclei, and understanding the many-body mechanisms involved in heavy nuclear fission. These areas will require theoretical, mathematical, and computational developments to make progress, and they will also require new facilities for experimental validation, such as the Facility for Radioactive Ion Beams, in order to provide validation of computational predictions. Figure 4 schematically indicates the problems to be addressed, and the computational requirements, measured in overall floating point operations (FLOPs) needs, to solve these problems.

**4. The role of computing in accelerating the innovation cycle**
In the previous section I discussed a particular area of nuclear physics that has benefitted significantly from the application of computing to solve the quantum many-body problem. The question remains – how can one export this kind of expertise into the private sector in order to enhance the innovation cycle? A recent OSTP report summarizes findings and describes strategies that could produce such an outcome [29].

One can ask first, has industry utilized high-performance computing, and has the use of computing helped a particular industry accelerate the design cycle? The answer is yes to both questions. Table II illustrates early impacts of what has been described as simulation-based engineering and science (SBES) [30]. In a number of these examples, a partnership with government enabled companies to accelerate innovation through application of computing and development of codes that effectively utilize available platforms. The government's role was to assist companies in that process. One of the DOE computing programs, Innovative and Novel Computational Impact on Theory and Experiment (INCITE), is open to non-proprietary industrial use of computing platforms, and has been somewhat successful in delivering results to various companies.

**Table 2: Early impacts of SBSS [31].**

| Company | Innovation | Impact |
|---|---|---|
| **Boeing** | Predictive optimization of airfoil design | 7-fold decrease in testing |
| **Cummins** | New engine brought to market solely with modelling and analysis tools | Reduced development time and cost; improved engine performance |
| **Goodyear** | Predictive modelling for new tire design | 3-fold reduction in product development time |
| **Ford** | Virtual aluminium casting | Estimated 7:1 return on investment; $100M in savings |
| **GE/P&W** | Accelerated insertion of materials in components | 50% reduction in development time, increased capability with reduced testing |

Several issues impede the general adoption of HPC in industry. Industries approach computing in terms of a business model: the value proposition for use of HPC is that businesses will increase innovation and decrease the development cycle. In many instances, particularly for smaller companies, that value proposition is not as clearly linked to product. Another issue that impedes broad acceptance of HPC in the market place is the need to understand uncertainty in calculations. As Doug Post and Larry Votta have stated [32], "The prediction challenge is now the most serious limiting factor for computational science." The current excitement over uncertainty quantification and a continuing push for validated codes flows from the prediction challenge. For example, a recent National Academy of Sciences (NAS) study recognized the value of Uncertainty quantification (UQ) for nuclear stockpile evaluations [33]. In a similar fashion, in order for a business to bet big on a computational output, there has to be substantial confidence in the results of the simulation.

Another important ingredient for the success and broader acceptance of the use of HPC in industry is a larger pool of educated scientists and engineers that are computationally aware. Universities should teach computational science and engineering to a broader set of students, including undergraduates. The goal is not that everyone should be able to write scientific simulations codes, but that students should be able to understand what makes a good and validated calculation. A few years ago, I taught a graduate-level course where the most common initial comment was, 'my code works because it compiles.' Industry needs people who have an in-depth understanding of what the results mean, and what the appropriate questions should be in order to utilize computational science to accelerate innovation.

Industrial use of computing today often relies on codes written several years ago. The barrier to changing a code involves validating that change experimentally, and also involves inertia. With computing capability in HPC continuing to double every two years, it is important that we develop relationships with industry that enables a fast development cycle for industrial codes as well. It is not that industrial applications need the largest systems (although sometimes they do), but they do need to understand how to utilize systems effectively. On node parallelism is going to be standard whether one utilizes a desktop or the largest machines, so there is already an impetus to update and rewrite codes.

It will be interesting to see how the story of computing and industrial competitiveness unfolds in the coming years. The challenges to implementation are several, but the rewards can be significant.

## 5. Conclusion

I will conclude by reiterating some of the points made throughout this article. The first point is that the global use of fossil fuels has generated an increasing amount of green-house gases and is leading to global climate change. The U.S. has an opportunity to lead the world in developing new technologies in order to reduce global reliance on fossil fuels. The second point is that the U.S. must lead in innovation and high-end manufacturing to be competitive in a global economy. This entails creating newer technologies with optimal designs and shorter design cycles. New technologies are not stand-alone, but need to be incorporated into complex systems where optimization for efficiency, reliability, and security is important. The third point is that simulations have become an important means toward acceleration of discovery in science and national security. The methodology has been developed that combines HPC with models and data in order to understand physical systems. The U.S. is a world leader in this capability, but other nations are moving quickly. The fourth point is that the for-profit sector has lagged in applying simulations to its problems. Only a handful of top companies have applied simulations to real-world problems, and the in-house capability in these companies is lagging. We have an opportunity to make simulation a unique asset for U.S. industry by transferring current capabilities, developing the workforce, and developing software and hardware that can be used to solve particular problems of interest. Furthermore, at the high end of computing there is much to do and we should pursue exascale computing in order to create that capability for both scientific discovery and national competitiveness [34].